\documentclass[showpacs,twocolumn,preprintnumbers]{revtex4-1} 
\usepackage{hyperref}
\pdfoutput=1
\usepackage{mathrsfs}
\usepackage{multirow}
\usepackage{tikz}
\usepackage{graphicx}
\usetikzlibrary{shapes.geometric, arrows}
\definecolor{light-gray}{gray}{0.95}

\newcommand{\C}{\mathbb{C}}

\newcommand{\cI}{\mathcal{I}}

\newcommand{\dbar}{\bar\partial}

\newcommand{\cM}{\mathcal{M}}

\newcommand{\be}{\begin{equation}\label}
\newcommand{\ee}{\end{equation}}
\newcommand{\bea}{\begin{eqnarray}\label}
\newcommand{\eea}{\end{eqnarray}}

\usepackage{graphicx,amsmath,amssymb}

\begin{document}
\preprint{CERN-TH-2016-172, DAMTP-2016-54}

\title{Two-Loop Scattering Amplitudes from the Riemann Sphere }

\author{Yvonne Geyer$^1$, Lionel Mason$^1$, Ricardo Monteiro$^2$, Piotr Tourkine$^3$ \vspace{.4cm}
\\ \small{$^1$Mathematical Institute, University of Oxford, Woodstock Road, Oxford OX2 6GG, UK
 \vspace{.1cm} \\
$^2$Theoretical Physics Department, CERN, Geneva, Switzerland
 \vspace{.1cm} \\
$^3$DAMTP, University of Cambridge, Wilberforce Road, Cambridge CB3 0WA, UK}}


\begin{abstract}
The scattering equations give striking formulae for massless scattering amplitudes at tree level and, as shown recently, at one loop. The progress at loop level was based on ambitwistor string theory, which naturally yields the scattering equations. We proposed that, for ambitwistor strings, the standard loop expansion in terms of the genus of the worldsheet is equivalent to an expansion in terms of nodes of a Riemann sphere, with the nodes carrying the loop momenta. In this paper, we show how to obtain two-loop scattering equations with the correct factorization properties.  We adapt genus-two integrands from the ambitwistor string to the nodal Riemann sphere and show that these yield correct answers, by matching standard results for the four-point two-loop amplitudes of maximal supergravity and super-Yang-Mills theory. In the Yang-Mills case, this requires the loop analogue of the Parke-Taylor factor carrying the colour dependence, which includes non-planar contributions.
\end{abstract}

\maketitle

\section{Introduction}
\label{sec:introduction}

The Cachazo-He-Yuan (CHY) formulae provide remarkable tree-level
expressions for scattering amplitudes in theories of massless
particles, written as an integral over marked points on the Riemann
sphere.  The integral localises as a sum over the solutions to the
scattering equations \cite{Cachazo:2013hca}. This formalism
generalizes earlier work of Roiban, Spradlin and Volovich
\cite{Roiban:2004yf} based on Witten's twistor string theory
\cite{Witten:2003nn}. The CHY formulae themselves originate in
ambitwistor string theory~\cite{Mason:2013sva}: this provided a
loop-level formulation~\cite{Adamo:2013tsa,Ohmori:2015sha} giving new
formulae at genus one (torus)~\cite{Adamo:2013tsa,Casali:2014hfa} and
two~\cite{Adamo:2015hoa} for type II supergravities in 10
dimensions. In~\cite{Geyer:2015bja,Geyer:2015jch}, we showed how the
torus formulae reduce to formulae on a nodal Riemann sphere, by means
of integration by parts in the moduli space of the torus. The node
carries the loop momentum. We proposed that an analogous reduction was
possible at any genus, leading to a new formalism that could become a
practical tool in the computation of scattering amplitudes. In the one-loop case, our explicit analysis provided a proof that the formulae from ambitwistor strings reproduce the correct answer. Furthermore, on the nodal Riemann sphere, the formalism is more flexible than on the torus, and the formulae could be extended to a variety of theories with or without supersymmetry. An alternative approach to the one-loop scattering equations was pursued in \cite{Cardona:2016bpi,Cardona:2016wcr}.

However, one loop is not such a stringent test of the framework, as many difficulties arise only at higher loops. The Feynman tree theorem, for example, shows how to construct one-loop integrands from tree formulae, if massive legs are allowed, and massive legs had already been considered in this context \cite{Naculich:2014naa}; an example of our formulae has been reproduced following such an approach \cite{Cachazo:2015aol}.  However, the situation is more difficult at higher loops despite recent progress inspired by the tree theorem \cite{Baadsgaard:2015twa}.

In \cite{Geyer:2015bja}, we gave a brief sketch as to how the loop-level scattering equations are obtained by reduction to the nodal Riemann sphere. In this Letter, we give a precise formulation at two loops. To fix the details of the reduction to the sphere, we use a factorization argument that leads to new
off-shell scattering equations.  An alternative approach   \cite{Feng:2016nrf}
 applies higher-dimensional tree-level rules for the integration of the scattering equations to give diagrams for a scalar theory; however, our aim here is to give a framework that yields loop integrands on a nodal Riemann sphere for complete amplitudes.  With this, we adapt genus-two supergravity integrands (type II, $d=10$) to a doubly nodal sphere, leading to the correct integrand for the four-point amplitude in maximal supergravity. We then conjecture an adjustment that gives instead a super-Yang-Mills integrand. These are checked both by factorization and numerically. Non-supersymmetric integrands require certain degenerate solutions to the scattering equations (on which the supersymmetric integrands vanish).  We characterize these degenerate solutions here, but leave the subtler non-supersymmetric integrands for the future.

\section{From higher genus to the sphere}
\label{sec:twoloopequations}
The higher genus scattering equations were formulated in the  ambitwistor-string
framework on Riemann surfaces $\Sigma_g$ of genus
$g$~\cite{Adamo:2013tsa,Ohmori:2015sha}, in terms of a meromorphic
1-form $P^\mu$, $\mu=1,\ldots,d$ 
(the momentum of the string) that solves
\begin{equation}
\bar\partial P=\sum_{i=1}^n k_i \,\delta^2( z -z_i) \,d z\wedge d\bar z,
\label{eq:X-path-int}
\end{equation}
where $ z _i$ are $n$ marked points on $ \Sigma _g$
.  The solution is written as
\begin{equation}
P= \sum_{i=1}^n k_i \omega_{z_i,z_0}^{(g)}(z)
+ \sum_{r=1}^g\ell_r
\omega_r^{(g)}\,,
\label{eq:P-loop}
\end{equation}
where $\omega_r^{(g)}$, $r=1,\ldots,g$ span a basis of holomorphic
1-forms on $\Sigma_g$ dual to a choice of a-cycles $a_r$, and
$\omega_{z_i,z_0}^{(g)}(z)$ are meromorphic differentials with simple
poles of residues $\pm1$ at $z_i$ and $z_0$ and vanishing $a$-cycle
integrals. The dependence on the auxiliary point $z_0$ drops by
momentum conservation.  The $\ell_r\in \mathbb C^d$ parametrize the
zero-modes of $P$ 
and will play the role of the loop
momenta.

The genus-$g$ scattering equations are a minimal set of conditions on the $z_i$ and moduli of $\Sigma_g$ required for $P^2$ to vanish globally. They include $n$ conditions
\begin{equation}
k_i\cdot P( z _i)=0 \, , \qquad i=1,\ldots , n\label{eq:P-res}
\end{equation}
that set the residues of the simple poles of $P^2$ at the $ z _i$ to zero.  These fix the locations of $ z _i$ on the surface. 
Once these are imposed, the quadratic differential $P^2$ is holomorphic, so it has $3g-3$ further degrees of freedom, corresponding to the moduli of the surface (its shape).  We therefore impose another $3g-3$ scattering equations to reach $P^2=0$.  
This can be done by writing
\begin{equation}
P^2= \sum_{r,s=1}^g u_{rs}\omega_r^{(g)}\omega^{(g)}_s\,,
\label{Psquare}
\end{equation}
where the $u_{rs}$ only depend on the 
moduli of $ \Sigma_g$ and on the kinematics, and setting an independent $3g-3$ subset of the $u_{rs}$ to zero. In total, the scattering equations localise the full moduli space integral to a discrete set of points. For $g=2,3$, there are precisely $3g-3$ $u_{rs}$'s, thus we simply set them all to zero and the ambitwistor-string loop integrand reads
\begin{equation}
\mathcal M_n^{(g)}=\int_{\cM_{g,n}} \mathcal I \;\mathrm d^{3g-3} \mu
\prod_{r\leq s} \bar{\delta}(u_{rs}) \prod_{i=1}^n \mathrm d  z _i \,
\bar{\delta}(k_i\cdot P(z_i))\label{eq:higher-genus} \,,
\end{equation}
where $\mathcal{I}$ is a correlator depending on the theory and the holomorphic $\delta$ functions are
$2\pi i \bar\delta(f(z))= \bar\partial({1}/f)$. We stress that this is a formula for the loop integrand, and the $n$-point $g$-loop amplitude is $\int d^D\ell_1\cdots d^D\ell_g\, {\mathcal M_n^{(g)}}$. 

To reduce this expression to one on nodal Riemann spheres, the heuristics described in \cite{Geyer:2015bja}, based on the explicit genus-one calculation, was to integrate by parts  (or use residue theorems) in the moduli integral $g$ times. This relaxes the delta functions $\dbar (1/u_{rr})$ to give measure factors $\prod_r 1/u_{rr}$, with the integration by parts yielding residues at the boundary of moduli space where all the chosen $a$-cycles contract to give double points, leaving a Riemann sphere with $g$ pairs of double points. This leaves $2g-3$ moduli that can be identified with the moduli of $2g$ points on the Riemann sphere, corresponding to the $g$ nodes, modulo M\"obius transformations.  These moduli 
are
fixed by $2g-3$ remaining scattering equations. 

On the nodal Riemann sphere $\Sigma$, the basis of 1-forms forms descending from the $\omega_r^{(g)}$'s dual to the pinched a-cycles, given by the pairs of double points $\sigma_{r^\pm}$, is
\begin{equation}
\omega_r=\frac{(\sigma_{r^+}-\sigma_{r^-}) d\sigma}{(\sigma-\sigma_{r^+})(\sigma-\sigma_{r^-})}\,, \quad r=1,\ldots,g
\label{eq:omega-sphere}
\end{equation}
so now
\begin{equation}
P=d\sigma\sum_{i=1}^n \frac{k_i}{\sigma-\sigma_i} +\sum_{r=1}^g
\ell_r\omega_r\, .\label{eq:P-sphere}
\end{equation}

From \eqref{Psquare}  the coefficient of the double poles at $\sigma_{r^\pm}$ identifies $u_{rr}$ as
$u_{rr}=\ell_r^2$.
Thus the measure factor becomes $\prod_r 1/\ell_r^2$.  Furthermore, the quadratic differential 
\begin{equation}
\mathcal{S}_g=P^2-\sum_{r=1}^g \ell_r^2 \omega_r^2\label{eq:S-def}
\end{equation}
now  only has simple poles at the $\sigma_i$ and $\sigma_{r^\pm}$. The $n+2g$  off-shell scattering equations were then proposed in  \cite{Geyer:2015bja} to be Res$_{\sigma_A}\mathcal{S}=0$.  There are three relations between these equations so that only $n+2g-3$ of them need to be imposed to enforce $\mathcal{S}=0$.  The three relations follow from the vanishing of the sum of residues of $\mathcal{S}$ multiplied by three independent tangent vectors to the sphere.

There is an ambiguity at two loops and higher, however.  We could equally well have defined $\mathcal{S}$ as
\begin{equation}
\widetilde{\mathcal{S}}_g=P^2-\sum_{r=1}^g \ell_r^2 \omega_r^2 + \sum_{r<s} a_{rs}
\omega_r\omega_s\label{eq:S-ambig} \,,
\end{equation}
where the $a_{rs}$ are linear combinations of the $u_{rs}$.  We  will see that $a_{rs}=\alpha(u_{rr}+u_{ss})=\alpha (\ell^2_r+\ell_s^2)$ is a better choice where $\alpha=\pm 1$ at two loops. This does not change the heuristic argument as it corresponds to  replacing the original $u_{rs}=0$ scattering equations for the moduli by nondegenerate linear combinations thereof.  The choice $\alpha=1$ (or equivalently -1) at two loops will be forced upon us by requiring correct factorisation channels. 

Thus our formula on the nodal Riemann sphere is
\begin{equation}
{\mathcal M}_n^{(g)}=\frac{1}{\prod_{r=1}^g \ell_r^2}\int_{\cM_{0,n+2g}} \hspace{-.2cm} \mathcal I_0 \frac{d^{n+2g}\sigma}{\mathrm{vol}\, SL(2,\C)}\prod_{A=1}^{n+2g}\hspace{-.1cm}{}^\prime\, \bar\delta(E_A) \,,
\label{eq:nodal-final}
\end{equation}
where 
$E_A = \mathrm{Res}_{\sigma_A}\widetilde{\mathcal{S}}_g$, with the
index $A$ spanning the $n$ marked points and  the $2g$ double points.
The delta functions enforce the off-shell scattering
equations~\footnote{with the $'$ on the product denoting the omission
  of three of the delta functions in line with the $SL(2,\C)$ quotient
  and the three relations between the $E_A$}.  In the first instance,
$\cI_0$ will be taken to be the nodal limit of the higher-genus
worldsheet correlator from the ambitwistor string type II
supergravity in $d=10$ (together with a cross ratio motivated by factorization). This can be extended to theories for which no higher-genus expression is known, as we will demonstrate explicitly for super-Yang-Mills theory.

\section{The 2-loop scattering equations}
We now take $g=2$ with  $\sigma_{1^\pm}$ and $\sigma_{2^\pm}$ the double points corresponding to $\ell_1$ and $\ell_2$. The two-loop scattering equations are the vanishing of the residues of
\begin{equation}
\mathcal{S}:=P^2- \ell_1^2 \omega_1^2 -\ell_2^2 \omega_2^2+ \alpha(\ell^2_1+\ell_2^2)
\omega_1\omega_2\, . \label{eq:S2}
\end{equation}
We adopt the shorthand notation $(AB)=\sigma_{A}-\sigma_{B}$. 
The scattering equations $2E_A=
\mathrm{Res}_{\sigma_A} S(\sigma)$ are then given by
\begin{align}\label{eq:SE-2loop}
\pm E_{1^\pm} & = \frac12\frac{L\,(2^+2^-)}{(1^\pm2^+)(1^\pm2^-)}
+ \sum_{i} \frac{\ell_1\cdot k_i}{(1^\pm i)}  \,, \nonumber\\
\pm E_{2^\pm} & =  \frac{1}{2} \, \frac{L\,(1^+1^-)}{(2^\pm 1^+)(2^\pm 1^-)} 
+ \sum_{i} \frac{\ell_2\cdot k_i}{(2^\pm i)}  \,,\\
E_i & = \frac{k_i\cdot\ell_1(1^+1^-)}{(i 1^+)(i 1^-)}
+ 
\frac{k_i\cdot\ell_2 (2^+2^-)}{(i2^+)(i2^-)} + \sum_{j\neq i} \frac{k_i\cdot k_j}{(ij)}\,,\nonumber
\end{align}
where $L=\alpha(\ell_1^2+\ell_2^2)+2\ell_1\cdot\ell_2$. In particular, for $\alpha=\pm 1$, $L=\pm(\ell_1\pm\ell_2)^2$.
The equations are not independent, since there are three linear relations between them,
\begin{equation}
\sum_A E_A=0 \,, \quad \sum_A \sigma_A E_A=0 \,, \quad \sum_A \sigma_A^2 E_A=0 \,.
\end{equation}
We will see that $\alpha=\pm1$ follows from  the correct factorisation.

\subsection{Poles and factorization}
Factorization channels of the integrand are related by the scattering equations to the boundary of the moduli space of the Riemann surface, where a subset of the marked points coalesce. Conformally, these configurations are equivalent to keeping the marked points at a finite distance, but pinching them off on another sphere, connected to the original one at the coalescence point $\sigma_I$. 

When $\Sigma$ degenerates in this way, the scattering equations force a kinematic configuration where an intermediate momentum goes on-shell  \cite{Dolan:2013isa}, corresponding to a potential pole in the integrand. The pole can thus be calculated as  $\sigma_A\rightarrow \sigma_I$ for  $A\in I$ from
\begin{equation}\label{eq:SE-poles}
\sum_{A\in I} (\sigma_A-\sigma_I)E_A=0\,.  
\end{equation}  
Note however that whether this singularity is realized in a specific theory depends on the integrand $\cI_0$. 

 Let $K_I=\sum_{i\in I} k_i$ (with external particles only). The location of the singularities in terms of the external and loop momenta can then be characterized as follows:
\begin{itemize}
 \item   When $\sigma_{1^{\pm}},\sigma_{2^{\pm}}\notin I$,  \eqref{eq:SE-poles} simply gives $K_I^2=0$, the standard factorization channel as for tree amplitudes, where a pole can appear in some intermediate propagator in massless scattering.  
 \item When $\sigma_{1^\pm}\in I$, but $\sigma_{2^{\pm}}\notin I$, we find the (potential) pole $2\ell_1\cdot K_I \pm  K_I^2$ as at 1-loop \cite{Geyer:2015bja}, where such poles arise from certain partial fraction relations and shifts in the loop momenta,  and coincide with the `Q-cut' poles \cite{Baadsgaard:2015twa}.  
\item The crucial new configuration at two loops is given by $\sigma_{1^+}, \sigma_{2^\pm}\in I$, corresponding to the condition
\begin{equation}
L + 2(\ell_1\pm \ell_2)\cdot K_I + K_I^2=0 \,,
\end{equation}
with $L$ as above. As detailed in \cite{Baadsgaard:2015twa}, the partial fraction identities and shifts always give a quadratic propagator of the form $(\ell_1\pm \ell_2+K_I)^2$ at two loops; see also Appendix~\ref{sec:shift-integr-plan}. Therefore, requiring the correct behaviour under factorisation determines $\pm=+$ and $\alpha=1$, or  $\pm=-$ and $\alpha=-1$. While both options are fully equivalent up to reparametrisation of the loop momenta, we will choose the former for the rest of the paper.

For  $\sigma_{1^+}, \sigma_{2^+}\in I$, this choice leads to a potential pole at $(\ell_1+ \ell_2+K_I)^2$. However, for $\sigma_{1^+}, \sigma_{2^-}\in I$, we are left with an unphysical potential pole at $(\ell_1+ \ell_2)^2+2(\ell_1-\ell_2)\cdot K_I +K_I^2=0$.
The requirement that this pole is absent from the final answer will give important restrictions on the integrand $\cI_0$.
\item Let us briefly comment on the only other new scenario at 2-loops -- to have both $\sigma_{1^\pm}\in I$. Since their contributions cancel in \eqref{eq:SE-poles}, this just leads to $K_I^2=0$, although now 
associated to two 1-loop diagrams joined by an on-shell propagator.
\end{itemize}

The criterion for an integrand $\cI_0$ to give a simple kinematic pole in the final formula at one of these potential singularities is that $\cI_0$ should have a pole of order $2|I|-2$ as the marked points coalesce, as described in detail in \S4.1 of \cite{Geyer:2015jch}. If the pole has lower degree, the final formula will not have a factorisation pole in this channel. This gives an important criterion for determining the precise forms of possible integrands $\cI_0$.

\subsection{Degenerate and regular solutions}
\begin{figure}[ht]
\centering
\input{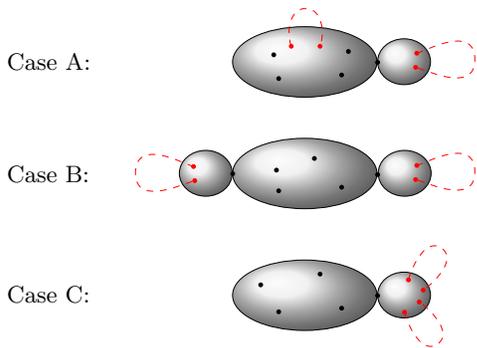}
\caption{Different possible worldsheet degenerations.}
\label{fig:degen}
\end{figure}

The off-shell scattering equations $\pm E_{r^\pm}(\sigma_{r^\pm})=0$ associated to the node $r$ have the same functional form, when seen as functions of  $\sigma=\sigma_{r^+}$ and $\sigma_{r^-}$ respectively. We distinguish between `regular solutions', when $\sigma_{r^{\pm}}$ are different roots of ${E}_{r^\pm}(\sigma)=0$, and `degenerate' solutions with $\sigma_{r^{+}}=\sigma_{r^-}$ the same root, for generic momenta.
These degenerate solutions can be summarized by the factorization diagrams given in figure \ref{fig:degen}, and can be understood as forward limits of the ($d+g$ dimensional) tree-level scattering equations; see Appendix \ref{sec:degen-sols} for details and \cite{He:2015yua,Cachazo:2015aol} for a discussion at one loop. 

However, not all solutions of the $d+2$ dimensional scattering equations survive in the forward limit. Consider a degeneration parameter $\tau$ which vanishes in the forward limit. While there are degenerate solutions with $\sigma_{r^+r^-}\sim\tau^2$, the zero-locus of ~\eqref{eq:SE-2loop} excludes them, and thus the two-loop integrands localise on the degenerate solutions with $\sigma_{r^+r^-}\sim\tau$ and on
\begin{equation}
  \label{eq:nb-sols-2loops}
  N_{\text{reg}}=(n+1)!-4n!+4(n-1)!+6(n-3)!
\end{equation}
regular solutions (with $\sigma_{r^+r^-}\sim1$). Moreover, we shall see that the supersymmetric integrands only receive contributions from the regular solutions.
As an important consequence, unphysical poles in the form of Gram determinants arising from double roots in $E_{r^\pm}(\sigma)$ are absent for supersymmetric theories: as discussed in \cite{Cachazo:2015aol} and appendix \ref{sec:degen-sols}, these poles can be localised on the degenerate solutions. For non-supersymmetric theories, however, degenerate solutions may contribute, and one must check that contributions with unphysical poles vanish upon loop integration, as detailed in \cite{Cachazo:2015aol} at one loop.

\section{Supersymmetric two-loop amplitudes}
\label{sec:genus-two}

We now consider explicit expressions at four points for maximal supergravity and super-Yang-Mills. These expressions are examples of \eqref{eq:nodal-final} for $g=2$, $n=4$. The representation of loop integrands that arises can be connected to a standard Feynman-like representation, after use of partial fractions and shifts in the loop momenta as  in \cite{Geyer:2015bja} at one-loop; see Appendix~\ref{sec:shift-integr-plan}.

\subsection{Four-point supergravity integrand}
We use  the integrand  that  arises directly from the degeneration of the genus-two
(ambitwistor) string
\cite{Adamo:2015hoa,D'Hoker:2002gw,Berkovits:2005df,Berkovits:2005ng}.
Define
\begin{equation}
  \Delta_{i,j}=\omega_1  (\sigma_i) \,\omega_2(\sigma_j)-\omega_1  (\sigma_j)\, \omega_2(\sigma_i)
\label{eq:Delta-def}
\end{equation}
and
\begin{equation}
{\hat{\mathcal Y}} = \frac{(k_1-k_2)\cdot (k_3-k_4)\, \Delta_{1,2}\Delta_{3,4} + \text{cyc(234)}}{3(1^+2^+)(1^+2^-)(1^-2^+)(1^-2^-)},
\label{eq:Yhat-def}
\end{equation}
where $\text{cyc(234)}$ is a sum over cyclic permutations.
 Our
prescription for the four-point supergravity integrand is
\begin{equation}
  \label{eq:4ptsugraSE}
 {\mathcal I}_0^{\text{SUGRA}} = ({\mathcal{K}}\tilde {\mathcal{K}}) \;{\hat{\mathcal Y}^2}\;
\frac{(1^+2^-)(1^-2^+)}{(1^+1^-)(2^+2^-)},
\end{equation}
where ${\mathcal{K}}\tilde {\mathcal{K}}$ is the standard kinematical supersymmetry prefactor ${\mathcal{K}}=(k_1\cdot k_2)(k_2\cdot k_3)A_{\text{tree}}^{\text{SYM}}(1,2,3,4)$ and ${\tilde{ \mathcal{K}}}=(k_1\cdot k_2)(k_2\cdot k_3){\tilde A}_{\text{tree}}^{\text{SYM}}(1,2,3,4)$, so that the supergravity states in the scattering are the direct product of super-Yang-Mills untilded ({\it left}) and tilded ({\it right}) states. 

The cross-ratio is inserted by hand to remove poles in the unphysical factorization channels discussed in the previous section. This is an important aspect of our prescription. We set the relative sign between $\ell_1$ and $\ell_2$ to be $+$, consistently with the $(\ell_1+\ell_2)^2$ factors in the scattering equations. This choice implies that the degenerations of the worldsheet at $(1^+2^+)\rightarrow 0$ or $(1^-2^-)\rightarrow 0$ occur when $(\ell_1+\ell_2\pm K_I)^2\rightarrow 0$, where $K_I$ is a partial sum of the external momenta. These physical poles can be realized in the formula. However, the numerator in the cross ratio suppresses unphysical poles of the type $(\ell_1+\ell_2)^2\pm 2(\ell_1-\ell_2)\cdot K_I+K_I^2 \rightarrow 0$, which might have arisen when $(1^+2^-)\rightarrow 0$ or $(1^-2^+)\rightarrow 0$.

We have evaluated our formula numerically and checked that it matches the known result for this amplitude \cite{Bern:1998ug},
\begin{align}
  \label{eq:2loopsugraint}
\frac{\mathcal{M}_4^{(2)}}{{\mathcal{K}}\tilde {\mathcal{K}}} =(k_1\cdot k_2)^2 \big[&  I^{\text{planar}}_{12,34} + I^{\text{planar}}_{34,21}
+ I^{\text{non-planar}}_{1,2,34} \nonumber \\
& + I^{\text{non-planar}}_{3,4,21} \big] + \text{cyc(234)}.
\end{align}
The planar and non-planar double-box integrands are written down in the ``shifted'' representation in Appendix~\ref{sec:shift-integr-plan}. It is also possible to check the factorization of this formula  explicitly.  

\subsection{Four-point super-Yang-Mills integrand}
There is no fully well defined ambitwistor model that would give a first principle
derivation of 
a super-Yang-Mills integrand,  however the tree and one-loop results
motivated us to postulate the following expression
\begin{equation}
  \label{eq:4ptsymSE}
 {\mathcal I}_0^{\text{SYM}} = {\mathcal{K}} \;{\hat{\mathcal Y}}\;
{\mathcal I}_{\text{PT}(2)},
\end{equation}
where ${\mathcal{K}}$ is again the standard kinematical supersymmetry prefactor. The new and crucial ingredient is the extension to two loops of the Park-Taylor factor, ${\mathcal I}_{\text{PT}(2)}$. In~\cite{Geyer:2015bja}, we presented the analogous object at one loop. We will comment later on the general form, and first focus on the explicit formula for the four-point two-loop case, including the non-planar (NP) contributions,
\begin{align}
  \label{eq:PT2loop}
{\mathcal I}_{\text{PT}(2)} =& \big[N_c^2 \,C^{\text{P}}_{1234}+ C^{\text{NP},1}_{1234}\big]
\big[ \text{tr}(1234)+\text{tr}(4321)\big] \nonumber \\
& + N_c\,C^{\text{NP},2}_{12,34} \,\text{tr}(12)\text{tr}(34) + \text{cyc(234)},
\end{align}
where $N_c$ is the rank of the gauge group and $\text{tr}(12\cdots)\equiv\text{tr}(T^{a_1}T^{a_2}\cdots)$ denote the colour traces. We have, for the planar part
\begin{align}
C^{\text{P}}_{1234} =& c^{\text{P}}_{1234}(1^+,1^-,2^+,2^-) +c^{\text{P}}_{1234}(1^-,1^+,2^-,2^+) \nonumber \\
& + c^{\text{P}}_{1234}(2^+,2^-,1^+,1^-) +c^{\text{P}}_{1234}(2^-,2^+,1^-,1^+), \nonumber
\end{align}
with
\begin{align}
c^{\text{P}}_{1234}(a,b,c,d) &= \frac{1}{(abdc1234)} + \frac{1}{(ab1dc234)} \nonumber \\
&\hspace{-.2cm} + \frac{1/2}{(ab12dc34)} + \frac{1}{(acdb1234)} + \text{cyc(1234)} \nonumber,
\end{align}
where $(123\cdots m)$ stands for $(12)(23)\cdots (m1)$. For the double-trace contribution, we have
\begin{align}
C^{\text{NP},2}_{12,34} =& c^{\text{NP}}_{12,34}(1^+,1^-,2^+,2^-) +c^{\text{NP}}_{12,34}(1^-,1^+,2^-,2^+) 
\nonumber \\
& + c^{\text{NP}}_{12,34}(2^+,2^-,1^+,1^-) +c^{\text{NP}}_{12,34}(2^-,2^+,1^-,1^+), \nonumber
\end{align}
with
\begin{align}
c^{\text{NP}}_{12,34}(a,b,c,d) =& \frac{1/2}{(ac12db34)} + \frac{2}{(acd12b34)} \nonumber \\
&  + \frac{1}{(a1cd2b34)} + \text{perm(12,34)} \nonumber,
\end{align}
where $\text{perm(12,34)}$ denotes the eight permutations 
$(1\leftrightarrow2)$, $(3\leftrightarrow4)$ and 
$(12)\leftrightarrow(34)$.
The remaining contribution is determined by the ones already given, as seen in~\cite{Naculich:2011ep},
\begin{equation}
C^{\text{NP},1}_{1234} = 2 \big[C^{\text{P}}_{1234}+C^{\text{P}}_{1342}+C^{\text{P}}_{1423}\big]-C^{\text{NP},2}_{13;24}.
\end{equation}

We checked numerically that our proposal matches the known result of~\cite{Bern:1997nh}. For instance, using the colour decomposition of eq.~\eqref{eq:PT2loop}, we get for the planar part
\begin{align}
  \label{eq:2loopsymint}
\frac{\mathcal{M}^{(2)\text{P}}_{1234}}{{\mathcal{K}}} =k_1\cdot k_2\; I^{\text{planar}}_{12,34} +k_4\cdot k_1\; I^{\text{planar}}_{41,23}.
\end{align}

The two-loop Parke-Taylor formula ${\mathcal I}_{\text{PT}(2)}$ is non-trivial, and may seem hard to extend for higher multiplicity or loop order. We propose, however, that in general it can be computed from the correlator of a current algebra on the Riemann sphere, which was our procedure at four points. This extends the tree-level result of \cite{Nair:1988bq} and, more generally, follows by analogy to the heterotic string~\cite{Gross:1984dd}, where gauge interactions have a closed string-like nature as in ambitwistor string theory~\cite{Mason:2013sva}. The sum over states at a node of the Riemann sphere translates into a sum over the Lie algebra index of two additional operator insertions per loop momentum. 
To eliminate the contributions from the unwanted poles, we drop Parke-Taylor terms that have orderings where loop momentum insertions appear with alternate signs as in $(1^+\cdot1^-\cdot2^+\cdot2^-\cdot)$, keeping only terms with orderings of the type $(1^+\cdot1^-\cdot2^-\cdot2^+\cdot)$; here $\cdot$ denote any external particles. For instance, we keep contributions such as $1/(1^+1^-2^-2^+1234)$, but discard terms like $1/(1^+1^-2^+2^-1234)$. This achieves the same effect as the cross-ratio appearing in the supergravity integrand. Moreover, we only include contributions with a single cyclic structure, e.g.~(123456), and discard contributions with subcycles, e.g.~(123)(456). These properties can be verified in the expressions above. Our two-loop Parke-Taylor expressions should be applicable to Yang-Mills theories with or without supersymmetry.

\section{Discussion}

We have obtained scattering equation formulae for two-loop integrands on the Riemann sphere, following the heuristic reduction of genus-two ambitwistor string formulae by integration by parts on the moduli space of Riemann surfaces as in \cite{Geyer:2015bja}. Our analysis is not a rigorous derivation from the genus-two ambitwistor string formulae of \cite{Adamo:2015hoa}, and in particular does not fix the parameter $\alpha$ in the off-shell scattering equations.   Nevertheless, we have seen that factorization  fixes the ambiguity in $\alpha $, and  this choice leads to correct two-loop integrands for maximally supersymmetric theories. A more refined analysis of the ambitwistor string degeneration should uniquely fix the scattering equations and the details of the integrands (such as the cross ratio in the supergravity case). It would also give us the tools to address the higher-loop case, where we expect different boundary contributions (e.g. at three loops ``mercedes'' vs. ``ladder'' graphs) in the integration by parts on the moduli space associated to different classes of scattering equations.

There are clearly many other challenges.
It should also be possible to obtain integrands for non-supersymmetric theories, as we did in \cite{Geyer:2015jch} at one loop.  These will in principle also have support on the degenerate solutions to the two-loop scattering equations, which we studied here; see \cite{Bjerrum-Bohr:2016juj,Bosma:2016ttj,Cardona:2016gon} and references therein for recent work on the scattering equations. For both supersymmetric and non-supersymmetric Yang-Mills and gravity at higher points, we need to understand  the higher-loop analogues of the CHY Pfaffians on the Riemann sphere with and without supersymmetry. The extent of supersymmetry should be determined by the particular sum over spin structures, as in \cite{Geyer:2015jch} at one loop. More generally, we would like to extend our results to a new formalism, where our formulae arise directly as correlation functions of vertex operators on the nodal Riemann sphere. A natural question is then what type of quantum field theories admit such a formulation;  there are CHY formulae and ambitwistor string models for a variety of theories \cite{Cachazo:2014xea,Ohmori:2015sha,Casali:2015vta}. Finally, it would be important to clarify the relation of these ideas to full string theory, which has been the subject of recent works \cite{Siegel:2015axg,Huang:2016bdd,Casali:2016atr}.


\section*{Acknowledgements}

We would like to thank Tim Adamo, Zvi Bern, Eduardo Casali and David Skinner for discussions. We also thank NORDITA, Stockholm, and the Isaac Newton Institute for Mathematical Sciences, Cambridge, for hospitality and financial support from  EPSRC grant EP/K032208/1 during the program GTA 2016. YG is supported by the EPSRC Doctoral Prize Scheme EP/M508111/1, LJM by the EPSRC grant EP/M018911/1, and the work of PT is supported by STFC grant ST/L000385/1.

\onecolumngrid
\appendix

\section{Shifted integrands for planar and non-planar double boxes}
\label{sec:shift-integr-plan}

In this section, we give the representation of double-box integrands with shifted loop momenta that appear in the formalism of the loop-level scattering equations \cite{Geyer:2015bja}. 
Consider the planar double-box integrand with standard quadratic propagators,
\begin{equation}
\frac{1}{\ell_1^2(\ell_1-k_1)^2(\ell_1+k_2)^2\ell_2^2(\ell_2-k_4)^2(\ell_2+k_3)^2(\ell_{12}+k_{23})^2},
\end{equation}
where we use the notation $\ell_{12}=\ell_1+\ell_2$ and $k_{ij}=k_i+k_j$.
The choice of canonical loop momenta arising from the shifts requires that this integrand is split into two different contributions -- one for which $\ell_1+\ell_2$ is in the middle of the box, ($A$), and one for which $\ell_1$ or $\ell_2$ is, ($B$). In the case ($A$), we obtain 9 terms, corresponding to applying the partial fraction identity and shifts to the 3 factors containing only $\ell_1$, and to the 3 factors containing only $\ell_2$. This gives
\begin{equation}
I^A_{12,34} (\ell_1,\ell_2)= \frac{1}{\ell_1^2\ell_2^2} V_1^aM_{ab}V_2^b,
\end{equation}
with
\begin{align}
V_1 = & \left( \frac{1}{(-2\ell_1\cdot k_1)(2\ell_1\cdot k_2)},
\frac{1}{(2\ell_1\cdot k_1)(2\ell_1\cdot k_{12}+k_{12}^2)},
\frac{1}{(-2\ell_1\cdot k_2)(-2\ell_1\cdot k_{12}+k_{12}^2)} \right),\nonumber \\ 
V_2 = & \left( \frac{1}{(-2\ell_2\cdot k_4)(2\ell_2\cdot k_3)},
\frac{1}{(2\ell_2\cdot k_4)(2\ell_2\cdot k_{34}+k_{34}^2)},
\frac{1}{(-2\ell_2\cdot k_3)(-2\ell_2\cdot k_{34}+k_{34}^2)} \right),\nonumber \\
M = &
\left( \begin{array}{ccc}
\frac{1}{(\ell_{12}+k_{23})^2} & \frac{1}{(\ell_{12}-k_{1})^2} & \frac{1}{(\ell_{12}+k_{2})^2} \\
\frac{1}{(\ell_{12}-k_{4})^2} & \frac{1}{\ell_{12}^2} & \frac{1}{(\ell_{12}+k_{12})^2} \\
\frac{1}{(\ell_{12}+k_{3})^2} & \frac{1}{(\ell_{12}+k_{34})^2} & \frac{1}{\ell_{12}^2} \end{array} \right). \nonumber
\end{align}
In case ($B$), take $\ell_1$ to be in the  middle propagator of the box first, as in
\begin{equation}
\frac{1}{\ell_1^2\ell_2^2(\ell_2-k_4)^2(\ell_2+k_3)^2(\ell_{12}+k_3)^2(\ell_{12}-k_4)^2(\ell_{12}+k_{23})^2},
\end{equation}
There are again 9 terms, because there are 3 factors with $\ell_2$ and 3 factors with $\ell_{12}$. After symmetrising also in the choice of $\ell_2$ versus $\ell_{12}$, we get
\begin{equation}
I^B_{12,34} (\ell_1,\ell_2)= \frac{1}{\ell_1^2\ell_2^2} V_{12\,a} V_2^a,
\end{equation}
with $V_2$ given as above and
\begin{align}
V_{12}= & \Big( v(\ell_{12}),v(\ell_{12}+k_4),v(\ell_{12}-k_3) \Big), \nonumber \\
v(\ell_{12}) = & \frac{1}{(\ell_{12}+k_{23})^2(-2\ell_{12}\cdot k_2-k_{23}^2)(2\ell_{12}\cdot k_1-k_{23}^2)}
\nonumber \\
&  + \frac{1}{(\ell_{12}+k_{3})^2(2\ell_{12}\cdot k_2+k_{23}^2)(-2\ell_{12}\cdot k_{34})} + \frac{1}{(\ell_{12}-k_{4})^2(-2\ell_{12}\cdot k_1+k_{23}^2)(2\ell_{12}\cdot k_{34})}. \nonumber
\end{align}
The total contribution from the planar double-box is obtained after the further symmetrisation of the loop momentum choices:
\begin{equation}
I^{\text{planar}}_{12,34} = \frac{1}{12}  \Big(I^A_{12,34} + I^B_{12,34} + I^B_{34,12} + [\ell_1\leftrightarrow \ell_2] + [\ell_1\to-\ell_1,\ell_2\to-\ell_2] \Big),
\end{equation}
where the numerical factor takes into account the symmetrisations over three types of shifts and the four types of loop momentum choices.

The non-planar double-box is analogous. Starting with case (C) where there are three propagators with $\ell_{12}$, 
\begin{equation}
\frac{1}{\ell_1^2(\ell_1-k_2)^2\ell_2^2(\ell_2-k_1)^2\ell_{12}^2(\ell_{12}+k_4)^2(\ell_{12}+k_{34})^2},
\end{equation}
we obtain
\begin{equation}
I^C_{1,2,34} (\ell_1,\ell_2)= \frac{1}{\ell_1^2\ell_2^2}\, \frac{1}{(2\ell_1\cdot k_2)(2\ell_2\cdot k_1)}\, \Big( t(\ell_{12})+t(\ell_{12}+k_{12})-t(\ell_{12}+k_1)-t(\ell_{12}+k_2) \Big),
\end{equation}
with
\begin{align}
t(\ell_{12}) =& \frac{1}{\ell_{12}^2(2\ell_{12}\cdot k_4)(2\ell_{12}\cdot k_{34}+k_{34}^2)}+
\frac{1}{(\ell_{12}+k_4)^2(-2\ell_{12}\cdot k_4)(2\ell_{12}\cdot k_3+k_{34}^2)} \nonumber \\
&+ \frac{1}{(\ell_{12}+k_{34})^2(-2\ell_{12}\cdot k_{34}-k_{34}^2)(-2\ell_{12}\cdot k_3-k_{34}^2)}. \nonumber
\end{align}
The other case, (D), is when there three propagators with $\ell_1$ or $\ell_2$, say $\ell_2$:
\begin{equation}
\frac{1}{\ell_1^2(\ell_1-k_2)^2\ell_2^2(\ell_2-k_4)^2(\ell_2-k_{34})^2\ell_{12}^2(\ell_{12}+k_1)^2}.
\end{equation}
Then we get
\begin{equation}
I^D_{1,2,34} (\ell_1,\ell_2)= \frac{1}{\ell_1^2\ell_2^2} U_1^aN_{ab}U_2^b,
\end{equation}
with
\begin{align}
U_1 = & \left( \frac{1}{(-2\ell_1\cdot k_2)},  \frac{1}{(2\ell_1\cdot k_2)} \right),\nonumber \\ 
U_2 = & \left( \frac{1}{(-2\ell_2\cdot k_4)(-2\ell_2\cdot k_{34}+k_{34}^2)},
\frac{1}{(2\ell_2\cdot k_4)(-2\ell_2\cdot k_3)},
\frac{1}{(2\ell_2\cdot k_{34}+k_{34}^2)(2\ell_2\cdot k_3)} \right),\nonumber \\
N = &
\left( \begin{array}{ccc}
u(\ell_{12}) & u(\ell_{12}+k_4) & u(\ell_{12}+k_{34}) \\
u(\ell_{12}+k_2) & u(\ell_{12}+k_{24}) & u(\ell_{12}-k_1) \end{array} \right),
\quad u(\ell_{12})=\left(\frac{1}{\ell_{12}^2}-\frac{1}{(\ell_{12}+k_1)^2}\right)\frac{1}{(2\ell_{12}\cdot k_1)^2}. \nonumber
\end{align}
Finally, the total contribution from the non-planar double-box is
\begin{equation}
I^{\text{non-planar}}_{1,2,34} = \frac{1}{12}  \Big(I^C_{12,34} + I^D_{12,34} + I^D_{34,12} + [\ell_1\leftrightarrow \ell_2] + [\ell_1\to-\ell_1,\ell_2\to-\ell_2] \Big).
\end{equation}

\section{Analysis of the degenerate solutions}
\label{sec:degen-sols}
In this section, we give a more detailed analysis of the two-loop scattering equations and their solutions. As at one loop \cite{He:2015yua,Cachazo:2015aol}, the key is to study the $d+g$ dimensional (massless) tree-level scattering equations for $2g$ additional particles, then reduce to $d$ dimensions and take the forward limit: 
\begin{equation}
 d\mu_{0,n+4}^{(d+2)}\rightarrow d\mu_{2,n}^{(d)}\equiv \frac{\prod_A\sigma_A}{\text{vol}(\text{SL}(2,\C))} \prod_A\bar\delta(E_A)\,.
\end{equation}
In particular, while this procedure reconstructs the two-loop scattering equations $E_A$ on the nodal Riemann sphere  ~\eqref{eq:SE-2loop}, it retains enough information of the massive scattering equations (in $d$ dimensions) to analyse the different classes of solutions. The main incentive for this study is an unphysical pole arising from double roots in the loop scattering equations. We will see explicitly how this pole can be reduced to a specific subset of the solutions, which do not contribute for the supersymmetric theories discussed in this Letter.

At two loops, our starting point are thus the $d+2$-dimensional massless scattering equations for $n+4$ particles with momenta $\{k_{1^\pm},k_{2^\pm},k_i\}$, 
\begin{align}\label{eq:SE_d+2}
 \mathcal{E}^{(d+2)}_A=\sum_{B}\frac{k_A\cdot k_B}{(AB)}\,,
\end{align}
with $A\in\{1^\pm,2^\pm,1,\dots,n\}$, where we have suggestively indexed the particles that will give rise to the loop momentum under the forward limit by $1^\pm$ and $2^\pm$. In particular, we take the external particles $k_i$ to only have components in $d$ dimensions, and we denote this $d$ dimensional part of $k_{1^\pm},k_{2^\pm}$ by $\tilde{\ell}_{1^\pm},\tilde{\ell}_{2^\pm}$ respectively. It is now always possible to choose the remaining components of $k_{1^\pm},k_{2^\pm}$ such that the scattering equations reduce to 
\begin{subequations}
\begin{align}
 \mathcal{E}^{(d)}_{1^\pm}&=\frac{\tilde{\ell}_{1^+}\cdot \tilde{\ell}_{1^-}+m_1^2}{(1^\pm1^\mp)}+\sum_{r=2^\pm}\frac{\tilde{\ell}_{1^\pm}\cdot \tilde{\ell}_{r}\pm\frac1{2}(m_1^2+m_2^2)}{(1^\pm r)}+\sum_{i=1}^n \frac{\tilde{\ell}_{1^\pm}\cdot k_i}{(1^\pm i)}\\
 \mathcal{E}^{(d)}_{2^\pm}&=\frac{\tilde{\ell}_{2^+}\cdot \tilde{\ell}_{2^-}+m_2^2}{(2^\pm2^\mp)}+\sum_{r=1^\pm}\frac{\tilde{\ell}_{2^\pm}\cdot \tilde{\ell}_{r}\pm\frac1{2}(m_1^2+m_2^2)}{(2^\pm r)}+\sum_{i=1}^n \frac{\tilde{\ell}_{2^\pm}\cdot k_i}{(2^\pm i)}\\
 \mathcal{E}^{(d)}_i&=\sum_{j=1}^n\frac{k_i\cdot k_j}{(ij)} +\sum_{r=1^\pm,2^\pm}\frac{k_i \cdot\tilde{\ell}_{r}}{(ir)}
\end{align}
\end{subequations}
Note in particular that in the forward limit (governed by a parameter $\tau\rightarrow 0$), where
\begin{align} \label{eq:forward-limit}
 \tilde{\ell}_{r^\pm}=\pm \ell_r +\frac{\tau}{2}q_r\,,\qquad \ell_r^2=m_r^2\,,
\end{align}
these equations smoothly limit onto the two-loop scattering equations ~\eqref{eq:SE-2loop}. However, as first pointed out in \cite{He:2015yua} at one loop, not all their solutions have a smooth limit as well -- the zero-locus of ~\eqref{eq:SE-2loop} excludes a subset of the solutions. To see this, first recall that we distinguish two different classes of solutions: since the two-loop scattering equations $E_r\equiv\pm E_{r^\pm}$ have the same functional form as functions of $\sigma=\sigma_{r^\pm}$ respectively, there are both `regular solutions' with $\sigma_{r^\pm}$ localising on different roots of $ E_r$ and `degenerate solutions', where $\sigma_{r^+}=\sigma_{r^-}$. Moreover, perturbing around the soft limit, the degenerate solutions come in three variations, see figure \ref{fig:degen}:
\begin{itemize}
 \item[] case A: \hspace{10pt} $\sigma_{1^+}=\sigma_{1^-}$, but $\sigma_{2^+}\neq\sigma_{2^-}$ (or $\sigma_{2^+}=\sigma_{2^-}$, but $\sigma_{1^+}\neq\sigma_{1^-}$)
 \item[] case B:  \hspace{10pt} $\sigma_{1^+}=\sigma_{1^-}$ and $\sigma_{2^+}=\sigma_{2^-}$, but $\sigma_{1^+}\neq\sigma_{2^+}$
 \item[] case C: \hspace{10pt} $\sigma_{1^+}=\sigma_{1^-}=\sigma_{2^+}=\sigma_{2^-}$.
\end{itemize}
For each degeneration of the nodal Riemann sphere, we distinguish furthermore between two types of solutions, depending on the rate of coalescence of $\sigma_{r^{\pm}}$. For the soft limit parameter $\tau$ as above, they behave as 
\begin{equation*}
\begin{aligned}
 &\text{regular:} && \sigma_{r^+r^-}\sim 1\\
 &\text{type I:} && \sigma_{r^+r^-}\sim\tau\\
 &\text{type II:} &&\sigma_{r^+r^-}\sim\tau^2\,.
\end{aligned}
\end{equation*}

While the type II solutions contribute for the $d+2$ dimensional tree-level scattering equations $\mathcal{E}^{(d+2)}_A$, the zero-locus of ~\eqref{eq:SE-2loop} excludes them, and thus the two-loop integrands localise on the type I solutions and the regular solutions.

\vspace{5pt}
{\bf Case A.}
To see this explicitly, let us perturb around the forward limit and focus on the case A. Both these solutions and the type B solutions bear a close resemblance to one-loop \cite{He:2015yua}, and our discussion will proceed in analogy. We take the forward limit ~\eqref{eq:forward-limit}, where $q$ is a fixed vector with $q^2\neq0$, and moreover
\begin{equation}
 \sigma_{2^{\pm}}=\sigma_I\pm\varepsilon +\mathcal{O}(\varepsilon^2)\,,
\end{equation}
to restrict to the degenerate solutions only, and then study the scattering equations perturbatively in $\varepsilon$ and $\tau$. To the relevant order, the scattering equations become
\begin{subequations}
\begin{align}
 \mathcal{E}_i^A&=\frac{\ell_1\cdot k_i}{(i1^+)}-\frac{\ell_1\cdot k_i}{(i1^-)}+\sum_{j=1}^n\frac{k_i\cdot k_j}{(ij)}+\mathcal{O}(\varepsilon,\tau)\,,\\
 \mathcal{E}_{1^\pm}^A&=\pm\sum_{i=1}^n \frac{\ell_{1}\cdot k_i}{(1^\pm i)}+\mathcal{O}(\varepsilon,\tau)\,,\\
 \mathcal{E}_{2^\pm}^A&=\pm\frac{\tau^2 q_2^2}{2\varepsilon}\pm\sum_{i=1}^n \frac{\ell_{2}\cdot k_i}{(Ii)}+\frac{\tau}{2}\sum_i\frac{q_2\cdot k_i}{(Ii)}-\varepsilon\sum_i\frac{\ell_2\cdot k_i}{(Ii)^2} +\\
 &\qquad\qquad +\frac{\pm \frac{1}{2}(\ell_1+ \ell_2)^2  +\frac{\tau}{2}(\pm q_1\cdot \ell_2 +q_2\cdot \ell_1)+\frac{\tau^2}{4}q_1\cdot q_2}{(I1^+)}+\nonumber\\
  &\qquad\qquad + \frac{\mp\frac{1}{2}(\ell_1+ \ell_2)^2  +\frac{\tau}{2}(\mp q_1\cdot \ell_2 -q_2\cdot \ell_1)+\frac{\tau^2}{4}q_1\cdot q_2}{(I1^-)} +\mathcal{O}(\varepsilon \tau, \varepsilon^2)\,.\nonumber
\end{align}
\end{subequations}
To leading order, the first two equations are the scattering equations at one loop, while the equations $\mathcal{E}_{2^\pm}$ are best understood in their polynomial form, $\mathcal{F}_0\equiv\mathcal{E}_{2^+}+\mathcal{E}_{2^-}$ and $\varepsilon \mathcal{F}_1\equiv\sigma_{2^+}\mathcal{E}_{2^+}+\sigma_{2^-}\mathcal{E}_{2^-}$;
\begin{subequations}
\begin{align}
 \mathcal{F}_0^A&=\tau \sum_{i=1}^n \frac{q_2\cdot k_i}{(Ii)}-\varepsilon\sum_{i=1}^n \frac{\ell_2\cdot k_i}{(Ii)^2}+\mathcal{O}(\varepsilon\tau, \tau^2)\label{eq:degen_caseA_1}\\
 \mathcal{F}_1^A&=\sum_{i=1}^n \frac{\ell_2\cdot k_i}{(Ii)}+\frac{\tau^2q_2^2}{\varepsilon}+\mathcal{O}(\tau, \varepsilon)\,.\label{eq:degen_caseA_2}
\end{align}
\end{subequations}
Evidently, there are two dominant balances for $\varepsilon$ in terms of $\tau$: $\varepsilon\sim\tau$ from ~\eqref{eq:degen_caseA_1} (type I in the notation above), and $\varepsilon\sim\tau^2$ from ~\eqref{eq:degen_caseA_2} (type II). We can now confirm that the zero-locus of the two-loop scattering equations excludes the type II solutions with $\varepsilon\sim\tau^2$, since this dominant balance includes a term $\tau^2q_2^2/\varepsilon\sim 1$, which is absent in ~\eqref{eq:SE-2loop}.

Moreover, each case gives $n$ solutions for the coalescence point $\sigma_I$, and one solution for $\varepsilon$. Together with the $(n-1)!-2(n-2)!$ (non-degenerate) solutions to the one-loop scattering equations on the nodal sphere, this leads to $n((n-1)!-2(n-2)!)$ solutions of each type. This is as expected: since the momentum flowing through the node connecting the two spheres is soft due to the back-to-back loop momenta $\pm \ell_2$, the location of the connecting node $\sigma_I$ does not affect the other $n+2$ marked points. The prefactor $n$ originates from the equation determining the location of the soft node in terms of the other marked points. This reflects the known results for a soft particle scattering with $N$ hard particles -- the scattering equations for the $N$ particles decouple, giving $(N-3)!$ solutions, while the equation determining the location of the soft particle is of degree $N-2$. 

\vspace{5pt}
{\bf Case B.} Case B follows by analogy. Starting from the forward limit and restricting to the degenerate solutions
\begin{equation}
 \sigma_{1^{\pm}}=\sigma_{I_1}\pm\varepsilon_1 +\mathcal{O}(\varepsilon_1^2)\,,\qquad\qquad\sigma_{2^{\pm}}=\sigma_{I_2}\pm\varepsilon_2 +\mathcal{O}(\varepsilon_2^2)\,,
\end{equation}
the tree-level scattering equations $\mathcal{E}_i$ decouple. To leading order, the remaining scattering equations in their polynomial form have the same functional form as in case A, $\mathcal{F}_{0,1}^A(\sigma_I, \varepsilon)=\mathcal{F}_{0,1}^B(\sigma_{I_2}, \varepsilon_2)=\mathcal{F}_{0,1}^B(\sigma_{I_1}, \varepsilon_1)|_{\ell_2\rightarrow\ell_1,q_2\rightarrow q_1}$. Therefore, there are four types of degenerate solutions, with $\sigma_{1^+1^-}\sim\sigma_{2^+2^-}\sim\tau$; $\sigma_{1^+1^-}\sim\tau$ and $\sigma_{2^+2^-}\sim\tau^2$ (and with $\sigma_{1^\pm}$, $\sigma_{2^\pm}$ interchanged); and $\sigma_{1^+1^-}\sim\sigma_{2^+2^-}\sim\tau^2$. However, only $\sigma_{1^+1^-}\sim\sigma_{2^+2^-}\sim\tau$ is a solution to the two-loop scattering equations for the same reason discussed above. When counting the number of solutions, there are now $(n-2)$ solutions for each of the coalescence points $\sigma_{I_{1,2}}$, and $(n-3)!$ solutions to the tree-level scattering equations, thus $(n-2)^2(n-3)!$ solutions for type I in total.

\vspace{5pt}
{\bf Case C.} While the cases A and B work in close analogy to one loop due to the decoupling of the scattering equations on the main Riemann sphere, the case C degeneration is a new feature at two loops. In contrast to the discussion above, there are now four remaining equations on the Riemann sphere containing the loop momenta, determining the location of the connecting node as well as the separations of the loop nodes.
The most convenient parametrisation for the marked points is
\begin{equation}\label{eq:caseC_degen}
 \sigma_{r^\pm}=\sigma_I+\varepsilon x_{r^\pm}\,,\qquad\qquad\text{with }x_{r^+}=x_r+y_r\,\text{ and }x_{r^-}=x_r\,.
\end{equation}
M\"{o}bius invariance on the sphere guarantees that we can always fix the locations of two of these, e.g. $x_r$ (with the connecting node taken to be at $x_I=\infty$ on the sphere containing the loop momenta). After imposing both the forward limit and ~\eqref{eq:caseC_degen} to restrict to case C, the scattering equations are most concise in the polynomial form
\begin{subequations}
\begin{align}
 \mathcal{F}_a^C=\sum_{r=1^\pm,2^\pm}x_r^a\mathcal{E}_r^C\,,\qquad\qquad a=0,1,2,3\,.
\end{align}
\end{subequations}
In particular, we find that $\mathcal{F}_3^C$ contains a term of the form
\begin{equation}
 \mathcal{F}_3^C=\frac{(\ell_1+\ell_2)^2}{2\varepsilon}y_1y_2+\mathcal{O}(1)\,,
\end{equation}
with all other terms of order $\mathcal{O}(1)$. Thus, we conclude that we only obtain solutions for
\begin{equation*}
 y_1\sim\tau\text{ and }y_2\sim 1\qquad\text{ or }\qquad
 y_1\sim 1\text{ and }y_2\sim \tau\qquad\text{ or }\qquad
 y_1\sim y_2\sim\tau\,.
\end{equation*}
A closer investigation reveals that in fact neither of the former cases gives a consistent dominant balance, and thus our solutions are of the form $y_1\sim y_2\sim\tau$,
\begin{equation}
 \sigma_{r^\pm}=\sigma_I+\varepsilon x_{r^\pm}\,,\qquad\qquad\text{with }x_{r^+}=x_r+\tau w_r\,\text{ and }x_{r^-}=x_r\,.
\end{equation}
This leads to only solutions of type I, with $\varepsilon\sim\tau$. Moreover, since $w_r$ and $\varepsilon$ appear linearly in the equations, there is exactly one solution for each of them. The remaining polynomial in $\sigma_I$ is of degree $6(2n-3)$ after eliminating $w_r$ and $\varepsilon$, and thus we find $6(2n-3)(n-3)!$ solutions.

\vspace{5pt}
{\bf Number of solutions.} This discussion can be summarised in table \ref{table:number_sol}. 
\begin{table}[t]
  \centering
  \begin{tabular}{|c|c|c|}
\hline
    {} &type I&type II \\ \hline
   case  A&$2 n ((n - 1)! - 2 (n - 2)!)$&$2 n ((n - 1)! - 2 (n - 2)!)$ \\ 
     case   B & $(n - 2)^2 (n - 3)!$ & $3(n - 2)^2 (n - 3)!$\\ 
    case C&$ 6 (2 n - 3) (n - 3)!$ & {}\\ \hline
  \end{tabular}
\caption{Number of solutions corresponding to various degenerations,
  as shown in fig.~\ref{fig:degen}}\label{table:number_sol}
\end{table}
As discussed, the type II solutions are excluded by the two-loop limit. The factor of 2 in the case A comes from interchanging $\ell_1$ and $\ell_2$. In particular, this implies that the two-loop scattering equations have $N_{\text{reg}}$ solutions, with
\begin{equation}
  N_{\mathrm{reg}}=(n+1)!-4n!+4(n-1)!+6(n-3)!\,.
\end{equation}

\vspace{5pt}
{\bf Degenerate solutions in supersymmetric theories.} Following the discussion in \S4.1 of \cite{Geyer:2015jch}, the measure for the degenerate solutions becomes
\begin{itemize}
 \item Case A: \hspace{10pt} $d\mu_{2,n}= \left(\bar\delta\left(\varepsilon-\tau F_A\right)\,d\varepsilon\,\,d\tilde{\mu}_I\right)\,d\mu_{1,n}$
 \item Case B: \hspace{10pt} $d\mu_{2,n}= \left(\bar\delta\left(\varepsilon_1-\tau F_{B_1}\right)\,d\varepsilon_1\,d\tilde{\mu}_{I_2}\right)\,\left(\bar\delta\left(\varepsilon_2-\tau F_{B_2}\right)\,d\varepsilon_2\,d\tilde{\mu}_{I_2}\right)\,d{\mu}_{0,n}$
 \item Case C: \hspace{10pt} $d\mu_{2,n}= \varepsilon^5\left(\bar\delta\left(\varepsilon-\tau F_C\right)\,d\varepsilon\,\,d\tilde{\mu}_I\right)\,d\mu_{0,n}$
\end{itemize}
Here, we have extracted the $\varepsilon$-dependence explicitly, so $d\tilde{\mu}_I$ represents the remaining measure on the loop-momentum Riemann sphere. For the super-Yang-Mills ~\eqref{eq:4ptsugraSE} and supergravity integrands ~\eqref{eq:4ptsymSE} at two loops, we find that
\begin{align*}
 &\text{Case A:}  && \mathcal{I}^{\text{SUGRA}}\sim \varepsilon^3 && \mathcal{I}^{\text{sYM}}\sim \varepsilon^1\\
 &\text{Case B:}  && \mathcal{I}^{\text{SUGRA}}\sim \varepsilon_1^3\varepsilon_2^3 && \mathcal{I}^{\text{sYM}}\sim \varepsilon_1^1\varepsilon_2^1\\
 &\text{Case C:}  && \mathcal{I}^{\text{SUGRA}}\sim 1 && \mathcal{I}^{\text{sYM}}\sim \varepsilon^{-2}\,.
\end{align*}
Therefore, all supersymmetric two-loop amplitudes behave as $\mathcal{O}(\varepsilon)$ for all degenerate solutions. The supersymmetric integrands thus only receive contributions from the regular solutions with $\sigma_{r^+r^-}\sim 1$.

\vspace{5pt}
{\bf Absence of the unphysical pole}
As pointed out in \cite{Cachazo:2015aol} at one loop, the loop-level scattering equations on the Riemann sphere contain unphysical poles. To see this, recall that the scattering equations at the nodes have the same functional form $E_{r}(\sigma)$, and the regular solutions are characterised by localising $\sigma_{r^\pm}$ on different roots. A special case thus arises when $E_{r}(\sigma)$ develops degenerate roots, and thus regular solutions become degenerate. As discussed above, when a solution is degenerate $\sigma_{r^+}=\sigma_{r^-}$, the scattering equations separate into lower loop-order equations, and the remaining marked points are independent of $\sigma_{r^\pm}$. Thus, denoting the sphere containing the marked points (and possibly one pair of nodes associated to the loop momenta in case A) by $\Sigma_I$, each unphysical pole is given by the discriminant of $E_r$ \cite{Cachazo:2015aol},
\begin{equation}
 \Delta=\prod_{\text{sol for }\sigma_i\in\Sigma_I}\text{Disc}(E_{r})\,.
\end{equation}
Note that this argument does not make any reference to the form of the other scattering equations (apart from requiring that they decouple for degenerate solutions), and thus the treatment of the unphysical pole at two loops proceeds exactly as at one loop. More specifically, integrating by parts localises the contribution of the unphysical pole to the degenerate solutions. In particular, since the integrands for super-Yang-Mills and supergravity vanish on these solutions, this implies that the degenerate pole never occurs.

For non-supersymmetric theories, more care is needed since the degenerate solutions will contribute in general. Just as at one loop, it is thus necessary to show that the contribution from the pole (and the integrand in this channel) is homogeneous in the loop momenta and thus vanishes upon integration.

\twocolumngrid

\bibliography{../twistor-bib}  

\begin{thebibliography}{35}%
\makeatletter
\providecommand \@ifxundefined [1]{%
 \@ifx{#1\undefined}
}%
\providecommand \@ifnum [1]{%
 \ifnum #1\expandafter \@firstoftwo
 \else \expandafter \@secondoftwo
 \fi
}%
\providecommand \@ifx [1]{%
 \ifx #1\expandafter \@firstoftwo
 \else \expandafter \@secondoftwo
 \fi
}%
\providecommand \natexlab [1]{#1}%
\providecommand \enquote  [1]{``#1''}%
\providecommand \bibnamefont  [1]{#1}%
\providecommand \bibfnamefont [1]{#1}%
\providecommand \citenamefont [1]{#1}%
\providecommand \href@noop [0]{\@secondoftwo}%
\providecommand \href [0]{\begingroup \@sanitize@url \@href}%
\providecommand \@href[1]{\@@startlink{#1}\@@href}%
\providecommand \@@href[1]{\endgroup#1\@@endlink}%
\providecommand \@sanitize@url [0]{\catcode `\\12\catcode `\$12\catcode
  `\&12\catcode `\#12\catcode `\^12\catcode `\_12\catcode `\%12\relax}%
\providecommand \@@startlink[1]{}%
\providecommand \@@endlink[0]{}%
\providecommand \url  [0]{\begingroup\@sanitize@url \@url }%
\providecommand \@url [1]{\endgroup\@href {#1}{\urlprefix }}%
\providecommand \urlprefix  [0]{URL }%
\providecommand \Eprint [0]{\href }%
\providecommand \doibase [0]{http://dx.doi.org/}%
\providecommand \selectlanguage [0]{\@gobble}%
\providecommand \bibinfo  [0]{\@secondoftwo}%
\providecommand \bibfield  [0]{\@secondoftwo}%
\providecommand \translation [1]{[#1]}%
\providecommand \BibitemOpen [0]{}%
\providecommand \bibitemStop [0]{}%
\providecommand \bibitemNoStop [0]{.\EOS\space}%
\providecommand \EOS [0]{\spacefactor3000\relax}%
\providecommand \BibitemShut  [1]{\csname bibitem#1\endcsname}%
\let\auto@bib@innerbib\@empty
\bibitem [{\citenamefont {Cachazo}\ \emph
  {et~al.}(2014{\natexlab{a}})\citenamefont {Cachazo}, \citenamefont {He},\
  and\ \citenamefont {Yuan}}]{Cachazo:2013hca}%
  \BibitemOpen
  \bibfield  {author} {\bibinfo {author} {\bibfnamefont {F.}~\bibnamefont
  {Cachazo}}, \bibinfo {author} {\bibfnamefont {S.}~\bibnamefont {He}}, \ and\
  \bibinfo {author} {\bibfnamefont {E.~Y.}\ \bibnamefont {Yuan}},\ }\href
  {\doibase 10.1103/PhysRevLett.113.171601} {\bibfield  {journal} {\bibinfo
  {journal} {Phys.Rev.Lett.}\ }\textbf {\bibinfo {volume} {113}},\ \bibinfo
  {pages} {171601} (\bibinfo {year} {2014}{\natexlab{a}})},\ \Eprint
  {http://arxiv.org/abs/1307.2199} {arXiv:1307.2199 [hep-th]} \BibitemShut
  {NoStop}%
\bibitem [{\citenamefont {Roiban}\ \emph {et~al.}(2004)\citenamefont {Roiban},
  \citenamefont {Spradlin},\ and\ \citenamefont {Volovich}}]{Roiban:2004yf}%
  \BibitemOpen
  \bibfield  {author} {\bibinfo {author} {\bibfnamefont {R.}~\bibnamefont
  {Roiban}}, \bibinfo {author} {\bibfnamefont {M.}~\bibnamefont {Spradlin}}, \
  and\ \bibinfo {author} {\bibfnamefont {A.}~\bibnamefont {Volovich}},\ }\href
  {\doibase 10.1103/PhysRevD.70.026009} {\bibfield  {journal} {\bibinfo
  {journal} {Phys.Rev.}\ }\textbf {\bibinfo {volume} {D70}},\ \bibinfo {pages}
  {026009} (\bibinfo {year} {2004})},\ \Eprint
  {http://arxiv.org/abs/hep-th/0403190} {arXiv:hep-th/0403190 [hep-th]}
  \BibitemShut {NoStop}%
\bibitem [{\citenamefont {Witten}(2004)}]{Witten:2003nn}%
  \BibitemOpen
  \bibfield  {author} {\bibinfo {author} {\bibfnamefont {E.}~\bibnamefont
  {Witten}},\ }\href {\doibase 10.1007/s00220-004-1187-3} {\bibfield  {journal}
  {\bibinfo  {journal} {Commun.Math.Phys.}\ }\textbf {\bibinfo {volume}
  {252}},\ \bibinfo {pages} {189} (\bibinfo {year} {2004})},\ \Eprint
  {http://arxiv.org/abs/hep-th/0312171} {arXiv:hep-th/0312171 [hep-th]}
  \BibitemShut {NoStop}%
\bibitem [{\citenamefont {Mason}\ and\ \citenamefont
  {Skinner}(2014)}]{Mason:2013sva}%
  \BibitemOpen
  \bibfield  {author} {\bibinfo {author} {\bibfnamefont {L.}~\bibnamefont
  {Mason}}\ and\ \bibinfo {author} {\bibfnamefont {D.}~\bibnamefont
  {Skinner}},\ }\href {\doibase 10.1007/JHEP07(2014)048} {\bibfield  {journal}
  {\bibinfo  {journal} {JHEP}\ }\textbf {\bibinfo {volume} {1407}},\ \bibinfo
  {pages} {048} (\bibinfo {year} {2014})},\ \Eprint
  {http://arxiv.org/abs/1311.2564} {arXiv:1311.2564 [hep-th]} \BibitemShut
  {NoStop}%
\bibitem [{\citenamefont {Adamo}\ \emph {et~al.}(2014)\citenamefont {Adamo},
  \citenamefont {Casali},\ and\ \citenamefont {Skinner}}]{Adamo:2013tsa}%
  \BibitemOpen
  \bibfield  {author} {\bibinfo {author} {\bibfnamefont {T.}~\bibnamefont
  {Adamo}}, \bibinfo {author} {\bibfnamefont {E.}~\bibnamefont {Casali}}, \
  and\ \bibinfo {author} {\bibfnamefont {D.}~\bibnamefont {Skinner}},\ }\href
  {\doibase 10.1007/JHEP04(2014)104} {\bibfield  {journal} {\bibinfo  {journal}
  {JHEP}\ }\textbf {\bibinfo {volume} {1404}},\ \bibinfo {pages} {104}
  (\bibinfo {year} {2014})},\ \Eprint {http://arxiv.org/abs/1312.3828}
  {arXiv:1312.3828 [hep-th]} \BibitemShut {NoStop}%
\bibitem [{\citenamefont {Ohmori}(2015)}]{Ohmori:2015sha}%
  \BibitemOpen
  \bibfield  {author} {\bibinfo {author} {\bibfnamefont {K.}~\bibnamefont
  {Ohmori}},\ }\href {\doibase 10.1007/JHEP06(2015)075} {\bibfield  {journal}
  {\bibinfo  {journal} {JHEP}\ }\textbf {\bibinfo {volume} {06}},\ \bibinfo
  {pages} {075} (\bibinfo {year} {2015})},\ \Eprint
  {http://arxiv.org/abs/1504.02675} {arXiv:1504.02675 [hep-th]} \BibitemShut
  {NoStop}%
\bibitem [{\citenamefont {Casali}\ and\ \citenamefont
  {Tourkine}(2015)}]{Casali:2014hfa}%
  \BibitemOpen
  \bibfield  {author} {\bibinfo {author} {\bibfnamefont {E.}~\bibnamefont
  {Casali}}\ and\ \bibinfo {author} {\bibfnamefont {P.}~\bibnamefont
  {Tourkine}},\ }\href {\doibase 10.1007/JHEP04(2015)013} {\bibfield  {journal}
  {\bibinfo  {journal} {JHEP}\ }\textbf {\bibinfo {volume} {1504}},\ \bibinfo
  {pages} {013} (\bibinfo {year} {2015})},\ \Eprint
  {http://arxiv.org/abs/1412.3787} {arXiv:1412.3787 [hep-th]} \BibitemShut
  {NoStop}%
\bibitem [{\citenamefont {Adamo}\ and\ \citenamefont
  {Casali}(2015)}]{Adamo:2015hoa}%
  \BibitemOpen
  \bibfield  {author} {\bibinfo {author} {\bibfnamefont {T.}~\bibnamefont
  {Adamo}}\ and\ \bibinfo {author} {\bibfnamefont {E.}~\bibnamefont {Casali}},\
  }\href {\doibase 10.1007/JHEP05(2015)120} {\bibfield  {journal} {\bibinfo
  {journal} {JHEP}\ }\textbf {\bibinfo {volume} {1505}},\ \bibinfo {pages}
  {120} (\bibinfo {year} {2015})},\ \Eprint {http://arxiv.org/abs/1502.06826}
  {arXiv:1502.06826 [hep-th]} \BibitemShut {NoStop}%
\bibitem [{\citenamefont {Geyer}\ \emph
  {et~al.}(2015{\natexlab{a}})\citenamefont {Geyer}, \citenamefont {Mason},
  \citenamefont {Monteiro},\ and\ \citenamefont {Tourkine}}]{Geyer:2015bja}%
  \BibitemOpen
  \bibfield  {author} {\bibinfo {author} {\bibfnamefont {Y.}~\bibnamefont
  {Geyer}}, \bibinfo {author} {\bibfnamefont {L.}~\bibnamefont {Mason}},
  \bibinfo {author} {\bibfnamefont {R.}~\bibnamefont {Monteiro}}, \ and\
  \bibinfo {author} {\bibfnamefont {P.}~\bibnamefont {Tourkine}},\ }\href
  {\doibase 10.1103/PhysRevLett.115.121603} {\bibfield  {journal} {\bibinfo
  {journal} {Phys. Rev. Lett.}\ }\textbf {\bibinfo {volume} {115}},\ \bibinfo
  {pages} {121603} (\bibinfo {year} {2015}{\natexlab{a}})},\ \Eprint
  {http://arxiv.org/abs/1507.00321} {arXiv:1507.00321 [hep-th]} \BibitemShut
  {NoStop}%
\bibitem [{\citenamefont {Geyer}\ \emph
  {et~al.}(2015{\natexlab{b}})\citenamefont {Geyer}, \citenamefont {Mason},
  \citenamefont {Monteiro},\ and\ \citenamefont {Tourkine}}]{Geyer:2015jch}%
  \BibitemOpen
  \bibfield  {author} {\bibinfo {author} {\bibfnamefont {Y.}~\bibnamefont
  {Geyer}}, \bibinfo {author} {\bibfnamefont {L.}~\bibnamefont {Mason}},
  \bibinfo {author} {\bibfnamefont {R.}~\bibnamefont {Monteiro}}, \ and\
  \bibinfo {author} {\bibfnamefont {P.}~\bibnamefont {Tourkine}},\ }\href@noop
  {} {\  (\bibinfo {year} {2015}{\natexlab{b}})},\ \Eprint
  {http://arxiv.org/abs/1511.06315} {arXiv:1511.06315 [hep-th]} \BibitemShut
  {NoStop}%
\bibitem [{\citenamefont {Cardona}\ and\ \citenamefont
  {Gomez}(2016{\natexlab{a}})}]{Cardona:2016bpi}%
  \BibitemOpen
  \bibfield  {author} {\bibinfo {author} {\bibfnamefont {C.}~\bibnamefont
  {Cardona}}\ and\ \bibinfo {author} {\bibfnamefont {H.}~\bibnamefont
  {Gomez}},\ }\href {\doibase 10.1007/JHEP06(2016)094} {\bibfield  {journal}
  {\bibinfo  {journal} {JHEP}\ }\textbf {\bibinfo {volume} {06}},\ \bibinfo
  {pages} {094} (\bibinfo {year} {2016}{\natexlab{a}})},\ \Eprint
  {http://arxiv.org/abs/1605.01446} {arXiv:1605.01446 [hep-th]} \BibitemShut
  {NoStop}%
\bibitem [{\citenamefont {Cardona}\ and\ \citenamefont
  {Gomez}(2016{\natexlab{b}})}]{Cardona:2016wcr}%
  \BibitemOpen
  \bibfield  {author} {\bibinfo {author} {\bibfnamefont {C.}~\bibnamefont
  {Cardona}}\ and\ \bibinfo {author} {\bibfnamefont {H.}~\bibnamefont
  {Gomez}},\ }\href@noop {} {\  (\bibinfo {year} {2016}{\natexlab{b}})},\
  \Eprint {http://arxiv.org/abs/1607.01871} {arXiv:1607.01871 [hep-th]}
  \BibitemShut {NoStop}%
\bibitem [{\citenamefont {Naculich}(2014)}]{Naculich:2014naa}%
  \BibitemOpen
  \bibfield  {author} {\bibinfo {author} {\bibfnamefont {S.~G.}\ \bibnamefont
  {Naculich}},\ }\href {\doibase 10.1007/JHEP09(2014)029} {\bibfield  {journal}
  {\bibinfo  {journal} {JHEP}\ }\textbf {\bibinfo {volume} {09}},\ \bibinfo
  {pages} {029} (\bibinfo {year} {2014})},\ \Eprint
  {http://arxiv.org/abs/1407.7836} {arXiv:1407.7836 [hep-th]} \BibitemShut
  {NoStop}%
\bibitem [{\citenamefont {Cachazo}\ \emph {et~al.}(2015)\citenamefont
  {Cachazo}, \citenamefont {He},\ and\ \citenamefont {Yuan}}]{Cachazo:2015aol}%
  \BibitemOpen
  \bibfield  {author} {\bibinfo {author} {\bibfnamefont {F.}~\bibnamefont
  {Cachazo}}, \bibinfo {author} {\bibfnamefont {S.}~\bibnamefont {He}}, \ and\
  \bibinfo {author} {\bibfnamefont {E.~Y.}\ \bibnamefont {Yuan}},\ }\href@noop
  {} {\  (\bibinfo {year} {2015})},\ \Eprint {http://arxiv.org/abs/1512.05001}
  {arXiv:1512.05001 [hep-th]} \BibitemShut {NoStop}%
\bibitem [{\citenamefont {Baadsgaard}\ \emph {et~al.}(2015)\citenamefont
  {Baadsgaard}, \citenamefont {Bjerrum-Bohr}, \citenamefont {Bourjaily},
  \citenamefont {Caron-Huot}, \citenamefont {Damgaard},\ and\ \citenamefont
  {Feng}}]{Baadsgaard:2015twa}%
  \BibitemOpen
  \bibfield  {author} {\bibinfo {author} {\bibfnamefont {C.}~\bibnamefont
  {Baadsgaard}}, \bibinfo {author} {\bibfnamefont {N.~E.~J.}\ \bibnamefont
  {Bjerrum-Bohr}}, \bibinfo {author} {\bibfnamefont {J.~L.}\ \bibnamefont
  {Bourjaily}}, \bibinfo {author} {\bibfnamefont {S.}~\bibnamefont
  {Caron-Huot}}, \bibinfo {author} {\bibfnamefont {P.~H.}\ \bibnamefont
  {Damgaard}}, \ and\ \bibinfo {author} {\bibfnamefont {B.}~\bibnamefont
  {Feng}},\ }\href@noop {} {\  (\bibinfo {year} {2015})},\ \Eprint
  {http://arxiv.org/abs/1509.02169} {arXiv:1509.02169 [hep-th]} \BibitemShut
  {NoStop}%
\bibitem [{\citenamefont {Feng}(2016)}]{Feng:2016nrf}%
  \BibitemOpen
  \bibfield  {author} {\bibinfo {author} {\bibfnamefont {B.}~\bibnamefont
  {Feng}},\ }\href {\doibase 10.1007/JHEP05(2016)061} {\bibfield  {journal}
  {\bibinfo  {journal} {JHEP}\ }\textbf {\bibinfo {volume} {05}},\ \bibinfo
  {pages} {061} (\bibinfo {year} {2016})},\ \Eprint
  {http://arxiv.org/abs/1601.05864} {arXiv:1601.05864 [hep-th]} \BibitemShut
  {NoStop}%
\bibitem [{Note1()}]{Note1}%
  \BibitemOpen
  \bibinfo {note} {With the $'$ on the product denoting the omission of three
  of the delta functions in line with the $SL(2,\protect \mathbb {C})$ quotient
  and the three relations between the $E_A$}\BibitemShut {NoStop}%
\bibitem [{\citenamefont {Dolan}\ and\ \citenamefont
  {Goddard}(2014)}]{Dolan:2013isa}%
  \BibitemOpen
  \bibfield  {author} {\bibinfo {author} {\bibfnamefont {L.}~\bibnamefont
  {Dolan}}\ and\ \bibinfo {author} {\bibfnamefont {P.}~\bibnamefont
  {Goddard}},\ }\href {\doibase 10.1007/JHEP05(2014)010} {\bibfield  {journal}
  {\bibinfo  {journal} {JHEP}\ }\textbf {\bibinfo {volume} {1405}},\ \bibinfo
  {pages} {010} (\bibinfo {year} {2014})},\ \Eprint
  {http://arxiv.org/abs/1311.5200} {arXiv:1311.5200 [hep-th]} \BibitemShut
  {NoStop}%
\bibitem [{\citenamefont {He}\ and\ \citenamefont {Yuan}(2015)}]{He:2015yua}%
  \BibitemOpen
  \bibfield  {author} {\bibinfo {author} {\bibfnamefont {S.}~\bibnamefont
  {He}}\ and\ \bibinfo {author} {\bibfnamefont {E.~Y.}\ \bibnamefont {Yuan}},\
  }\href@noop {} {\  (\bibinfo {year} {2015})},\ \Eprint
  {http://arxiv.org/abs/1508.06027} {arXiv:1508.06027 [hep-th]} \BibitemShut
  {NoStop}%
\bibitem [{\citenamefont {D'Hoker}\ and\ \citenamefont
  {Phong}(2002)}]{D'Hoker:2002gw}%
  \BibitemOpen
  \bibfield  {author} {\bibinfo {author} {\bibfnamefont {E.}~\bibnamefont
  {D'Hoker}}\ and\ \bibinfo {author} {\bibfnamefont {D.~H.}\ \bibnamefont
  {Phong}},\ }\bibfield  {booktitle} {\emph {\bibinfo {booktitle} {{Superstring
  theory. Proceedings, International Conference, Hangzhou, P.R. China, August
  12-15, 2002}}},\ }\href@noop {} {\bibfield  {journal} {\bibinfo  {journal}
  {Conf. Proc.}\ }\textbf {\bibinfo {volume} {C0208124}},\ \bibinfo {pages}
  {85} (\bibinfo {year} {2002})},\ \bibinfo {note} {[,85(2002)]},\ \Eprint
  {http://arxiv.org/abs/hep-th/0211111} {arXiv:hep-th/0211111 [hep-th]}
  \BibitemShut {NoStop}%
\bibitem [{\citenamefont {Berkovits}(2006)}]{Berkovits:2005df}%
  \BibitemOpen
  \bibfield  {author} {\bibinfo {author} {\bibfnamefont {N.}~\bibnamefont
  {Berkovits}},\ }\href {\doibase 10.1088/1126-6708/2006/01/005} {\bibfield
  {journal} {\bibinfo  {journal} {JHEP}\ }\textbf {\bibinfo {volume} {01}},\
  \bibinfo {pages} {005} (\bibinfo {year} {2006})},\ \Eprint
  {http://arxiv.org/abs/hep-th/0503197} {arXiv:hep-th/0503197 [hep-th]}
  \BibitemShut {NoStop}%
\bibitem [{\citenamefont {Berkovits}\ and\ \citenamefont
  {Mafra}(2006)}]{Berkovits:2005ng}%
  \BibitemOpen
  \bibfield  {author} {\bibinfo {author} {\bibfnamefont {N.}~\bibnamefont
  {Berkovits}}\ and\ \bibinfo {author} {\bibfnamefont {C.~R.}\ \bibnamefont
  {Mafra}},\ }\href {\doibase 10.1103/PhysRevLett.96.011602} {\bibfield
  {journal} {\bibinfo  {journal} {Phys. Rev. Lett.}\ }\textbf {\bibinfo
  {volume} {96}},\ \bibinfo {pages} {011602} (\bibinfo {year} {2006})},\
  \Eprint {http://arxiv.org/abs/hep-th/0509234} {arXiv:hep-th/0509234 [hep-th]}
  \BibitemShut {NoStop}%
\bibitem [{\citenamefont {Bern}\ \emph {et~al.}(1998)\citenamefont {Bern},
  \citenamefont {Dixon}, \citenamefont {Dunbar}, \citenamefont {Perelstein},\
  and\ \citenamefont {Rozowsky}}]{Bern:1998ug}%
  \BibitemOpen
  \bibfield  {author} {\bibinfo {author} {\bibfnamefont {Z.}~\bibnamefont
  {Bern}}, \bibinfo {author} {\bibfnamefont {L.~J.}\ \bibnamefont {Dixon}},
  \bibinfo {author} {\bibfnamefont {D.~C.}\ \bibnamefont {Dunbar}}, \bibinfo
  {author} {\bibfnamefont {M.}~\bibnamefont {Perelstein}}, \ and\ \bibinfo
  {author} {\bibfnamefont {J.~S.}\ \bibnamefont {Rozowsky}},\ }\href {\doibase
  10.1016/S0550-3213(98)00420-9} {\bibfield  {journal} {\bibinfo  {journal}
  {Nucl. Phys.}\ }\textbf {\bibinfo {volume} {B530}},\ \bibinfo {pages} {401}
  (\bibinfo {year} {1998})},\ \Eprint {http://arxiv.org/abs/hep-th/9802162}
  {arXiv:hep-th/9802162 [hep-th]} \BibitemShut {NoStop}%
\bibitem [{\citenamefont {Naculich}(2012)}]{Naculich:2011ep}%
  \BibitemOpen
  \bibfield  {author} {\bibinfo {author} {\bibfnamefont {S.~G.}\ \bibnamefont
  {Naculich}},\ }\href {\doibase 10.1016/j.physletb.2011.12.010} {\bibfield
  {journal} {\bibinfo  {journal} {Phys. Lett.}\ }\textbf {\bibinfo {volume}
  {B707}},\ \bibinfo {pages} {191} (\bibinfo {year} {2012})},\ \Eprint
  {http://arxiv.org/abs/1110.1859} {arXiv:1110.1859 [hep-th]} \BibitemShut
  {NoStop}%
\bibitem [{\citenamefont {Bern}\ \emph {et~al.}(1997)\citenamefont {Bern},
  \citenamefont {Rozowsky},\ and\ \citenamefont {Yan}}]{Bern:1997nh}%
  \BibitemOpen
  \bibfield  {author} {\bibinfo {author} {\bibfnamefont {Z.}~\bibnamefont
  {Bern}}, \bibinfo {author} {\bibfnamefont {J.~S.}\ \bibnamefont {Rozowsky}},
  \ and\ \bibinfo {author} {\bibfnamefont {B.}~\bibnamefont {Yan}},\ }\href
  {\doibase 10.1016/S0370-2693(97)00413-9} {\bibfield  {journal} {\bibinfo
  {journal} {Phys. Lett.}\ }\textbf {\bibinfo {volume} {B401}},\ \bibinfo
  {pages} {273} (\bibinfo {year} {1997})},\ \Eprint
  {http://arxiv.org/abs/hep-ph/9702424} {arXiv:hep-ph/9702424 [hep-ph]}
  \BibitemShut {NoStop}%
\bibitem [{\citenamefont {Nair}(1988)}]{Nair:1988bq}%
  \BibitemOpen
  \bibfield  {author} {\bibinfo {author} {\bibfnamefont {V.~P.}\ \bibnamefont
  {Nair}},\ }\href {\doibase 10.1016/0370-2693(88)91471-2} {\bibfield
  {journal} {\bibinfo  {journal} {Phys. Lett.}\ }\textbf {\bibinfo {volume}
  {B214}},\ \bibinfo {pages} {215} (\bibinfo {year} {1988})}\BibitemShut
  {NoStop}%
\bibitem [{\citenamefont {Gross}\ \emph {et~al.}(1985)\citenamefont {Gross},
  \citenamefont {Harvey}, \citenamefont {Martinec},\ and\ \citenamefont
  {Rohm}}]{Gross:1984dd}%
  \BibitemOpen
  \bibfield  {author} {\bibinfo {author} {\bibfnamefont {D.~J.}\ \bibnamefont
  {Gross}}, \bibinfo {author} {\bibfnamefont {J.~A.}\ \bibnamefont {Harvey}},
  \bibinfo {author} {\bibfnamefont {E.~J.}\ \bibnamefont {Martinec}}, \ and\
  \bibinfo {author} {\bibfnamefont {R.}~\bibnamefont {Rohm}},\ }\bibfield
  {booktitle} {\emph {\bibinfo {booktitle} {{In *Argonne/Chicago 1985,
  Proceedings, Anomalies, Geometry, Topology*, 299-313 and Preprint - Gross,
  D.J. (84,Rec.Jan.85) 11P}}},\ }\href {\doibase 10.1103/PhysRevLett.54.502}
  {\bibfield  {journal} {\bibinfo  {journal} {Phys. Rev. Lett.}\ }\textbf
  {\bibinfo {volume} {54}},\ \bibinfo {pages} {502} (\bibinfo {year}
  {1985})}\BibitemShut {NoStop}%
\bibitem [{\citenamefont {Bjerrum-Bohr}\ \emph {et~al.}(2016)\citenamefont
  {Bjerrum-Bohr}, \citenamefont {Bourjaily}, \citenamefont {Damgaard},\ and\
  \citenamefont {Feng}}]{Bjerrum-Bohr:2016juj}%
  \BibitemOpen
  \bibfield  {author} {\bibinfo {author} {\bibfnamefont {N.~E.~J.}\
  \bibnamefont {Bjerrum-Bohr}}, \bibinfo {author} {\bibfnamefont {J.~L.}\
  \bibnamefont {Bourjaily}}, \bibinfo {author} {\bibfnamefont {P.~H.}\
  \bibnamefont {Damgaard}}, \ and\ \bibinfo {author} {\bibfnamefont
  {B.}~\bibnamefont {Feng}},\ }\href@noop {} {\  (\bibinfo {year} {2016})},\
  \Eprint {http://arxiv.org/abs/1605.06501} {arXiv:1605.06501 [hep-th]}
  \BibitemShut {NoStop}%
\bibitem [{\citenamefont {Bosma}\ \emph {et~al.}(2016)\citenamefont {Bosma},
  \citenamefont {Sogaard},\ and\ \citenamefont {Zhang}}]{Bosma:2016ttj}%
  \BibitemOpen
  \bibfield  {author} {\bibinfo {author} {\bibfnamefont {J.}~\bibnamefont
  {Bosma}}, \bibinfo {author} {\bibfnamefont {M.}~\bibnamefont {Sogaard}}, \
  and\ \bibinfo {author} {\bibfnamefont {Y.}~\bibnamefont {Zhang}},\
  }\href@noop {} {\  (\bibinfo {year} {2016})},\ \Eprint
  {http://arxiv.org/abs/1605.08431} {arXiv:1605.08431 [hep-th]} \BibitemShut
  {NoStop}%
\bibitem [{\citenamefont {Cardona}\ \emph {et~al.}(2016)\citenamefont
  {Cardona}, \citenamefont {Feng}, \citenamefont {Gomez},\ and\ \citenamefont
  {Huang}}]{Cardona:2016gon}%
  \BibitemOpen
  \bibfield  {author} {\bibinfo {author} {\bibfnamefont {C.}~\bibnamefont
  {Cardona}}, \bibinfo {author} {\bibfnamefont {B.}~\bibnamefont {Feng}},
  \bibinfo {author} {\bibfnamefont {H.}~\bibnamefont {Gomez}}, \ and\ \bibinfo
  {author} {\bibfnamefont {R.}~\bibnamefont {Huang}},\ }\href@noop {} {\
  (\bibinfo {year} {2016})},\ \Eprint {http://arxiv.org/abs/1606.00670}
  {arXiv:1606.00670 [hep-th]} \BibitemShut {NoStop}%
\bibitem [{\citenamefont {Cachazo}\ \emph
  {et~al.}(2014{\natexlab{b}})\citenamefont {Cachazo}, \citenamefont {He},\
  and\ \citenamefont {Yuan}}]{Cachazo:2014xea}%
  \BibitemOpen
  \bibfield  {author} {\bibinfo {author} {\bibfnamefont {F.}~\bibnamefont
  {Cachazo}}, \bibinfo {author} {\bibfnamefont {S.}~\bibnamefont {He}}, \ and\
  \bibinfo {author} {\bibfnamefont {E.~Y.}\ \bibnamefont {Yuan}},\ }\href@noop
  {} {\  (\bibinfo {year} {2014}{\natexlab{b}})},\ \Eprint
  {http://arxiv.org/abs/1412.3479} {arXiv:1412.3479 [hep-th]} \BibitemShut
  {NoStop}%
\bibitem [{\citenamefont {Casali}\ \emph {et~al.}(2015)\citenamefont {Casali},
  \citenamefont {Geyer}, \citenamefont {Mason}, \citenamefont {Monteiro},\ and\
  \citenamefont {Roehrig}}]{Casali:2015vta}%
  \BibitemOpen
  \bibfield  {author} {\bibinfo {author} {\bibfnamefont {E.}~\bibnamefont
  {Casali}}, \bibinfo {author} {\bibfnamefont {Y.}~\bibnamefont {Geyer}},
  \bibinfo {author} {\bibfnamefont {L.}~\bibnamefont {Mason}}, \bibinfo
  {author} {\bibfnamefont {R.}~\bibnamefont {Monteiro}}, \ and\ \bibinfo
  {author} {\bibfnamefont {K.~A.}\ \bibnamefont {Roehrig}},\ }\href@noop {} {\
  (\bibinfo {year} {2015})},\ \Eprint {http://arxiv.org/abs/1506.08771}
  {arXiv:1506.08771 [hep-th]} \BibitemShut {NoStop}%
\bibitem [{\citenamefont {Siegel}(2015)}]{Siegel:2015axg}%
  \BibitemOpen
  \bibfield  {author} {\bibinfo {author} {\bibfnamefont {W.}~\bibnamefont
  {Siegel}},\ }\href@noop {} {\  (\bibinfo {year} {2015})},\ \Eprint
  {http://arxiv.org/abs/1512.02569} {arXiv:1512.02569 [hep-th]} \BibitemShut
  {NoStop}%
\bibitem [{\citenamefont {Huang}\ \emph {et~al.}(2016)\citenamefont {Huang},
  \citenamefont {Siegel},\ and\ \citenamefont {Yuan}}]{Huang:2016bdd}%
  \BibitemOpen
  \bibfield  {author} {\bibinfo {author} {\bibfnamefont {Y.-t.}\ \bibnamefont
  {Huang}}, \bibinfo {author} {\bibfnamefont {W.}~\bibnamefont {Siegel}}, \
  and\ \bibinfo {author} {\bibfnamefont {E.~Y.}\ \bibnamefont {Yuan}},\
  }\href@noop {} {\  (\bibinfo {year} {2016})},\ \Eprint
  {http://arxiv.org/abs/1603.02588} {arXiv:1603.02588 [hep-th]} \BibitemShut
  {NoStop}%
\bibitem [{\citenamefont {Casali}\ and\ \citenamefont
  {Tourkine}(2016)}]{Casali:2016atr}%
  \BibitemOpen
  \bibfield  {author} {\bibinfo {author} {\bibfnamefont {E.}~\bibnamefont
  {Casali}}\ and\ \bibinfo {author} {\bibfnamefont {P.}~\bibnamefont
  {Tourkine}},\ }\href@noop {} {\  (\bibinfo {year} {2016})},\ \Eprint
  {http://arxiv.org/abs/1606.05636} {arXiv:1606.05636 [hep-th]} \BibitemShut
  {NoStop}%
\end{thebibliography}%

\end{document}